\documentclass[newstyle,twocolumn,proceedings]{rmaa}
\usepackage{rmaacite}
\renewcommand{\P}[1]{%
\ifnum#1=1\hbox{OW~168--326E}\fi
\ifnum#1=2\hbox{OW~167--317}\fi
\ifnum#1=3\hbox{OW~163--317}\fi
\ifnum#1=5\hbox{OW~158--323}\fi
\ifnum#1=0\hbox{OW~171--334}\fi}

\newcommand{\beq}{\begin{equation}}
\newcommand{\eeq}{\end{equation}}

\newcommand{\kms}{\mbox{ km s$^{-1}$}~}

\newcommand{\mso}{\, M_\odot}

\newcommand{\rso}{\mbox{R$_{\odot}$}}

\newcommand{\cmj}{\, {\rm cm^2}\, {\rm s}^{-1}}

\newcommand{\Mo}{\mbox{M$_{\odot}$}~}

\title{Morphology and Galactic Distribution of PNs: \\ a New Scenario}
\author{G. Garc\'{\i}a-Segura, J. Franco, J. A. L\'opez,
N. Langer\altaffilmark{1} and M. R\'o\.zyczka\altaffilmark{2}
  \affil{Instituto de Astronom\'{\i}a-UNAM } }
\altaffiltext{1}{Astronomical Institute, Utrecht University, The Netherlands.}
\altaffiltext{2}{N. Copernicus Astronomical Center, Warszawa, Poland.}

\fulladdresses{Guillermo Garc\'{\i}a-Segura, 
Instituto de Astronom\'{\i}a-UNAM, Apdo Postal 877, 22800
Ensenada, B. C., Mexico }

\shortauthor{Garc\'{\i}a-Segura et al.}
\shorttitle{Morphology and Galactic Distribution of PNe}


\abstract{We review recent studies on the morphology and distribution
with respect to the galactic plane of planetary nebulae (PNs), as well
as recent advances in MHD modeling of PNs. We discuss a tentative
explanation for the connection between morphological classes and their
galactic distribution.}

\resumen{Se revisan trabajos sobre la morfologia y distribuci\'on
gal\'actica de las nebulosas planetarias (PNs), as\'{\i} como los
avances recientes en modelaje MHD de las PNs. Llegamos a una
explicaci\'on tentativa sobre la conexi\'on entre las clases
morfol\'ogicas y su distribuci\'on gal\'actica.}

\listofauthors{G. Garc\'{\i}a-Segura, J. Franco, J. A. L\'opez,
N. Langer \& M. R\'o\.zyczka}
\indexauthor{Garc\'{\i}a-Segura, G. }
\indexauthor{Franco, J.}
\indexauthor{L\'opez, J. A.}
\indexauthor{Langer , N.}
\indexauthor{R\'o\.zyczka, M.}

\begin{document}

\maketitle

\section{Introduction} 
The connection between PN morphologies and their distribution with
respect to the midplane of the Galaxy has been studied, with notorious
advances, during the last three decades. The search for systematic
segregations among PNs of different shapes started with the analysis
by Greig (1972). He had two main morphological classes, binebulous
(50\%) and circular (50\%), and found that the binebulous class had
a higher strength in the forbidden lines [O III], [O II] and [N II].
He also found a hint of a correlation between this class and smaller
distances to the galactic plane on average. Later, using the abundance
of chemical elements, Peimbert (P), Torres-Peimbert (T-P) and
collaborators (PT-P71, T-PP77, P78, T-PP79, PT-P83, T-PP83; see T-PP97
for a review) made a PN classification with four classes, from Type I
to Type IV, with decreasing abundances of Helium and heavy elements.
They found that most Type I PNs have bipolar shape, and this lead to
the conclusion that there might be population differences between
bipolar and circular PNs in our Galaxy. In the abstract of their
seminal paper, Calvet and Peimbert (1983) wrote: ``{\it It is
suggested that the bipolar nature of PN of Type I can be explained in
terms of their relatively massive progenitors ($ M \geq 2.4 $ \Mo),
that had to lose an appreciable fraction of their mass and angular
momentum during their planetary nebula stage. In a first mass-loss
stage at low velocity it is produced a disk and in a second mass-loss
stage at higher velocity the matter is limited by the disk, giving
rise to the bipolar structure}''. 

Later, several authors made improvements in our knowledge of this
issue. Zukerman \& Aller (1986) and Zukerman \& Gatley (1988)
classified 108 PNs in bipolar, round, disk-like, and annular. They
found an anti-correlation between metallicity and Galactic latitude,
as well as a segregation of morphological types according with
Galactic latitude, in agreement with previous studies. 

Balick (1987) made a major contribution to the morphological
classification (50 PNs), and named the classes as round, elliptical
and butterfly. At the same time, Chu et al.(1987) (126 PNs) published
a catalog of PNs with more than one shell (50 \% of the sample), and
named this new class as multiple-shell PNs. 

During the last decade, two important surveys were carried out on both
hemispheres. For the southern hemisphere, {\em The ESO Survey}, the
study was published in a series of papers by Schwarz, Corradi \&
Melnick (1992), Stanghellini, Corradi \& Schwarz (1993) (250 PNs),
Corradi \& Schwarz (1995), and Corradi (2000) (400 PNs). For the
northern hemisphere, {\em The IAC Survey}, the study was published by
Manchado et al.(1996) (243 PNs) and Manchado et al.(2000) (255 PNs). 
The first detailed study of the differences between ellipticals and
bipolars was done by Corradi \& Schwarz (1995) with the
ESO Survey (see also Corradi 2000). They found that the bipolar class
has a smaller scale height, 130 pc, than the one for ellipticals, 320
pc. Also, bipolars have the hottest central stars among PNe, and
display smaller deviations from pure circular Galactic rotation than
other morphological types. In addition, bipolars also display the
largest physical dimensions and have expansion velocities of up to an
order of magnitude above the typical values for PNe. These properties,
together with the chemical abundance results by Calvet \& Peimbert (1983),
indicate that bipolar PNe are produced by more massive progenitors
than the remaining morphological classes.

The distributions inferred by these surveys represent the main 
motivation for the present paper, which discusses these results in the
light of the new MHD models.
 
\section{Bipolars: Rotation and Angular Momentum} 

If bipolarity truly comes from the stellar rotation and the loss of 
angular momentum in massive PN progenitors, one can learn from the
case of the most massive stars and their associated nebulae. In the
Hertzsprung-Russell diagram, LBV stars like $\eta$~Car lie close to
the upper limit in temperature and luminosity beyond which no normal
stars are observed (the Humphreys-Davidson Limit), near the location
of the Eddington limit.
 
A massive star expanding at a constant luminosity $L$ would ---{\it in
the non-rotating case }--- reach the Eddington limit due to an opacity
increase in the surface layers. The Eddington limit is reached when
$\Gamma = L/L_{\rm Edd} = 1$, where the Eddington luminosity is 
$L_{\rm Edd}=4\pi cGM/\kappa$, with $c$ the speed of light, $M$ and
$L$ the mass and luminosity of the star, and $\kappa$ the opacity at
the stellar surface. Note that when $\Gamma \rightarrow 1$, the escape
velocity $v_{\rm esc}=[(1-\Gamma ) 2 G M / R]^{1/2} \rightarrow 0 $.
This implies a very slow wind velocity ($\simeq 0$) at the Eddington
limit. Thus, when a star passes through a close approach to the 
Eddington limit, its stellar wind will follow a {\it
fast---slow---fast} sequence. If we now include rotation, the critical
rotational velocity also follows the same behavior. The combination of
radiation pressure and centrifugal force exceeds surface gravity at
the equator when $v_{\rm crit}=[(1-\Gamma ) G M / R]^{1/2}$, and
$v_{\rm crit} \rightarrow 0$ when $\Gamma \rightarrow 1$. 
However, given that the rotational velocity $v_{\rm rot}$ can be small
but is always different from zero, critical rotation occurs {\it
before} the Eddington limit is reached. This is the so called $\Omega$
{\it limit} (Langer 1997). Winds with mass loss rates large enough to
halt and reverse the stellar expansion are thought to occur when
$\Omega=v_{\rm rot}/v_{\rm crit}$ approaches unity (Friend \& Abbott
1986). Thus, rotationally driven outflows will always occur before the
star reaches its Eddington limit.
 
Using this idea, Langer et al. (1999a) simulated the LBV outburst 
phenomenon by allowing the star to pass through three successive 
stages --- pre-outburst, outburst and post-outburst --- with
correspondingly different stellar winds. To include the effects of
stellar rotation on the winds, they used the analytic model of
Bjorkman \& Cassinelli (1993), with stellar parameters appropriate to
$\eta$~Car. The models were not intended to exactly fit all properties
of the Homunculus nebula. It is however striking how well these models,
without much fine tuning, not only reproduce the large-scale, bipolar
morphology, but also the small scale turbulent structure seen in
high-resolution observations of the Homunculus. This empirical result,
together with the observational facts that all LBV nebulae are bipolar
(Nota \& Clampin 1997), strongly indicates a link between rotation
and bipolarity.

{\it What about Bipolar PNs?}
In order to evaluate whether or not AGB rotation rates can produce
situations similar to the one discussed above, one needs to derive
$\Gamma$ and the rate of critical rotation. Direct determinations of
rotation rates for AGB stars are rare and mostly provide upper limits,
of the order of a few $\kms$ (except for V~Hydrae with $v\sin i\simeq
13\kms$). Using the available information for main sequence stars in
the initial mass range associated with PN progenitors (from $\sim
0.8\mso$ to about $5\mso$) these can be divided into two groups. Stars with
masses below $\sim 1.3\mso$, with subphotospheric convective zones,
spin down during their main sequence evolution due to flares and
magnetized winds. When they reach the AGB phase, their resulting
rotation velocities are below $0.01\kms$. Thus, the expected rotation
speeds of PN progenitors with masses below $\sim 1.3\mso$) should be
very small.

Stars above $\sim 1.3\mso$ do not have convective envelopes during core 
hydrogen burning, and they appear to remain as rapid rotators throughout 
their main sequence evolution. When 
they develop surface convection on their way to the red giant branch, they 
create a hydrogen burning shell which separates
the helium core from the envelope.  
The helium core evolves decoupled from the envelope and retains its
angular momentum, i.e. the entropy barrier of a nuclear burning 
shell prevents that angular momentum can leak out of the core.
Also in this case, however, the angular momentum 
of the hydrogen envelope will be lost, 
either due to magnetic braking or to mass 
loss and reexpansion of the convective envelope on the AGB
(cf. Langer et al.1999b).
As these
stars move to the thermally pulsing AGB phase, the H and He burning 
shells periodically switch on and off. Thus, the presumable barriers
vanish periodically and core-envelope angular momentum exchange can
occur during this stage. Such an exchange is very likely, since mixing
of matter through the core boundary is known to occur in this phase,
which is necessary to activate the $^{13}$C$(\alpha,n)$ reaction,
which is the neutron source to operate the s-process. To estimate the
resulting rotational velocity of the envelope (of about $0.1\mso$) at
the end of the AGB phase, we may approximate the specific angular
momentum of the core by its main sequence value, which is transferred
to the envelope. For a core of $\sim 0.5\mso$ and a main sequence 
radius of about $0.1\rso$ we get  $j\simeq 10^{17} ... 10^{18} \cmj$.
Adopting an average radius of the AGB envelope of $100\rso$, we obtain
a rotation velocity of $v_{rot}\simeq 10^{17.5}\cmj \times 0.5\mso /
100\rso / 0.1\mso \simeq 2\kms $. Regardless of how crude this
estimate may be, the net result is that stars at the tip of the RGB or
AGB phases may be subject to a significant spin-up of their surface
layers (Heger \& Langer 1998). Therefore, we propose here that values
of $\Omega= v_{rot}/v_{crit}$ close to 1 may be appropriate for AGB
single stars above $\sim 1.3\mso$ during the phase of PN ejection. We
parametrize the ``superwind'' in the same way as the giant LBV
outbursts (Eddington parameter $\Gamma=L/L_{\rm edd}$ close to~1; see
Garc\'{\i}a-Segura et al. 1999 for further details). 

There are variations of this scenario and the possibility exists that
stars which are not able to reach the $\Omega$ limit on their own
might be able to do so in the presence of a companion. For example,
main sequence wide binary systems, which become close binaries (either
detached or attached) at the AGB phase, can be subject to a very
effective spin-up by their companion. Depending on the binary
separation and masses, the
spin-up can be produced by tidal forces (slow process)
or by spiral-in (relatively fast process). In either cases,
the orbital angular momentum is transfered to the star.

Soker (1995) has studied the tidal spin-up applied to the
formation of elliptical, but the theory can be also applied
to the formation of bipolars (Soker \& Rappaport 2000).
Extreme cases of the same scenario is the 
common envelope evolution scenario (see for example Livio \&
Pringle 1996 and references there in, also Reyes-Ruiz \& L\'opez 1998).

Thus, a binary system of such a class can form a bipolar
nebula, however, the nebula will show point-symmetric
features, as we will discuss later on sections 4 and 5. 
Also, these nebulae do not have to show an enrichment
by heavy elements, and so, they do not necessary fit
into the Type I Peimbert category. Finally, since
their primary stars may be below $\sim 1.3\mso$ , their
scale height over the plane can be 
much larger than the one for classical Type I Peimbert nebulae
(see section 5).

\section{Ellipticals: A MHD Phenomenon} 

Axisymmetric flows can be produced by a magnetized wind with or
without the existence of equatorial density enhancements (EDE) (see
R\'o\.zyczka \& Franco 1996, Garc\'{\i}a-Segura 1997,
Garc\'{\i}a-Segura et al. 1999 and Garc\'{\i}a-Segura \& L\'opez
2000). An EDE can be formed with a small amount of rotation at the AGB
phase for example (Ignace et al.1998) or by a dipolar magnetic field
(Matt et al.2000). The magnetic field at the surface of a post-AGB
star can be transported out by its wind, similar but not equal to the
solar case. Because of stellar rotation, the magnetic field in the
wind is dominated by a toroidal component. The resulting toroidal
field has a magnetic tension associated with it. Thus, the general
effect of the  magnetic tension is the elongation of the nebula in the
polar direction. The mechanism responsible  for the elongation is
described by R\'o\.zyczka \& Franco (1996) and Franco et al. (2001,
this conference) in detail, and successful examples of the formation
of jets and ansae (FLIERS) can 
be seen in all above articles at the beginning of this section. 

Unfortunately, there is no information about stellar magnetic fields 
at the surfaces of post-AGB stars, which is in one of the most
interesting regions of the HR diagram, between the AGB phase and the
white dwarf phase. Reid et al.(1979) found strong circular
polarization in the OH masers from U Ori and IRC +10420. The OH masers
indicate fields of about 10 milligauss emanating from regions of about
$10^{15}$ cm from the star. They suggest that the field strengths at
the stellar surfaces are on the order of 10 and 100 Gauss,
respectively. White dwarfs are known to have fields of up to $10^7 -
10^8$ Gauss (Schmidt 1989). So, stellar field intensities between the
wide range defined by the solar value ($B_{\rm s}=2$ Gauss, in
average) and $10^8$ Gauss are acceptable for modeling. Note that
observations for the detection of magnetic fields in the swept-up
shells of PNe (eg Terzian 1989) have been tried, but however, they
give us very little information about the field intensity of the
nucleus, since that diluted field (if detected) will correspond to the
already expanded and compressed AGB gas.
 
\section{The Point-Symmetric Sub-Class: Signatures of Binaries} 

A particularly intriguing case in PNs morphologies are
those that display point-symmetric structures. 
At first view, the point-symmetric 
morphological class does not look very important. But, a careful inspection
to the statistically, complete sample of the IAC morphological catalog 
(Manchado et al.1996) reveals 40 objects with some degree of point-symmetric 
features, which represent  19\% of the total list (215) with a well 
defined morphology.
Note that many of these nebulae are not
classified as point-symmetric. In fact, they have been
well classified inside the categories of bipolars and ellipticals, 
such as the case of the Dumbel nebula. This fact points in the direction that
point-symmetry is a common feature related to any morphological
class, instead of a separate group (Manchado et al.2000; see also 
Guerrero, V\'azquez \& L\'opez 1998). 
Such a large fraction of the sample (19\%) suggest that 
the reason which produces point-symmetry should be very common indeed.

The most convincing solutions up to now, for the formation of 
point-symmetric nebulae require the existence of a binary system, 
and a magneto-hydrodynamical collimation of the wind,
either for accretion disk winds (Livio \& Pringle 1996; see also
Mastrodemos \& Morris 1998, 
Reyes-Ruiz \& L\'opez 1998 and Blackman et al. 2001)
or for stellar winds (Garc\'{\i}a-Segura 1997, Garc\'{\i}a-Segura \& 
L\'opez 2000).

Since the toroidal magnetic field carried out by the wind, either stellar or
coming from a disk (for the last one see Contopoulos 1995), is always 
perpendicular to the rotation
axis of the central star/disk (see Figure 3 in 
Garc\'{\i}a-Segura et al.2000), 
any kind of misalignment from the axis (wobbling instability, precession,
steady tilt respect to the equatorial density enhancement) 
will produce ``naturally'' a point-symmetric nebula.
In either case, it is easy to imagine the topology of the
magnetic field lines in such  scenarios, i.e., multiple
rings centered and aligned along the ``time-dependent'' 
spin axis of the star/disk.

Close binaries are a necessary condition in Livio \& Pringle (1996) , while
wide binaries are a sufficient one in Garc\'{\i}a-Segura (1997) and  
Garc\'{\i}a-Segura \& L\'opez (2000).
As described in section 2, close binaries, either attached or detached, 
will be in favor of forming bipolars with point-symmetric features,
while wide binaries will be in favor of ellipticals with 
point-symmetric features.
 
\section{Discussion} 

\begin{table}[!b]
\setlength{\tabcolsep}{2em}
\begin{center}
\caption{}
\label{}
\begin{tabular}{lc}
\hline
\hline
Morphological Class &  $< z >$    \\
 $\,$               &    pc       \\
\hline
B    &  110 \\
BPS  &  248 \\
E    &  308 \\
EPS  &  310 \\
R    &  753 \\
\hline
\hline
\end{tabular}
\end{center}
\end{table}
 
The average scale height over the plane, $< z >$, for different PN
morphological classes can be compared with those of different stellar
masses and populations. For instance, we know that massive stars are
located much closer to the galactic plane than the population of stars
with lower initial mass. The outcome from the ESO and IAC surveys is
certainly coincident: the bipolar (B) class has $< z >$ = 130 pc (ESO)
and $< z >$ = 179 pc (IAC), for ellipticals (E) $< z >$ = 320 pc (ESO)
and $< z >$ = 308 pc (IAC), and for rounds (R) $< z >$ = 753 pc (IAC).

In the recent analysis of the IAC Survey by Manchado et al. (2000),
the bipolar (BPS) and elliptical (EPS) objects with point-symmetric 
features were separated from those which do not present such a kind of
symmetries, i.e., from the B and E classes respectively. 
The new results from the IAC Survey are given in the Table 1.

Comparing the results of the IAC Survey with those described by Miller \&
Scalo (1979) for the average scale heigth of stars with different
masses, an average value equal or smaller than 110 pc corresponds to
stars with initial masses above 1.9 \Mo. Lower mass stars have average
scale heiths well above this value. These results are in line with the
pioneering suggestion of Calvet \& Peimbert (1983) and the more recent
discussion made by Garc\'{\i}a-Segura et al.(1999). 

To summarize, from the new MHD studies (sections 2, 3 and 4) 
we can conclude that:

\begin{description}

\item Bipolarity  $\Leftrightarrow$ Omega Limit $\Leftrightarrow$
 Angular Momentum (either orbital or stellar)

\item Ellipticity $\Leftrightarrow$ MHD effects $\Leftrightarrow$
 Symmetric Microstructures (FLIERS, jets,...etc)

\item Point-Symmetry $\Leftrightarrow$ Binaries  $\Leftrightarrow$
 Precession, Wobbling, ...

\end{description}

\begin{figure}
  \includegraphics[width=\columnwidth]{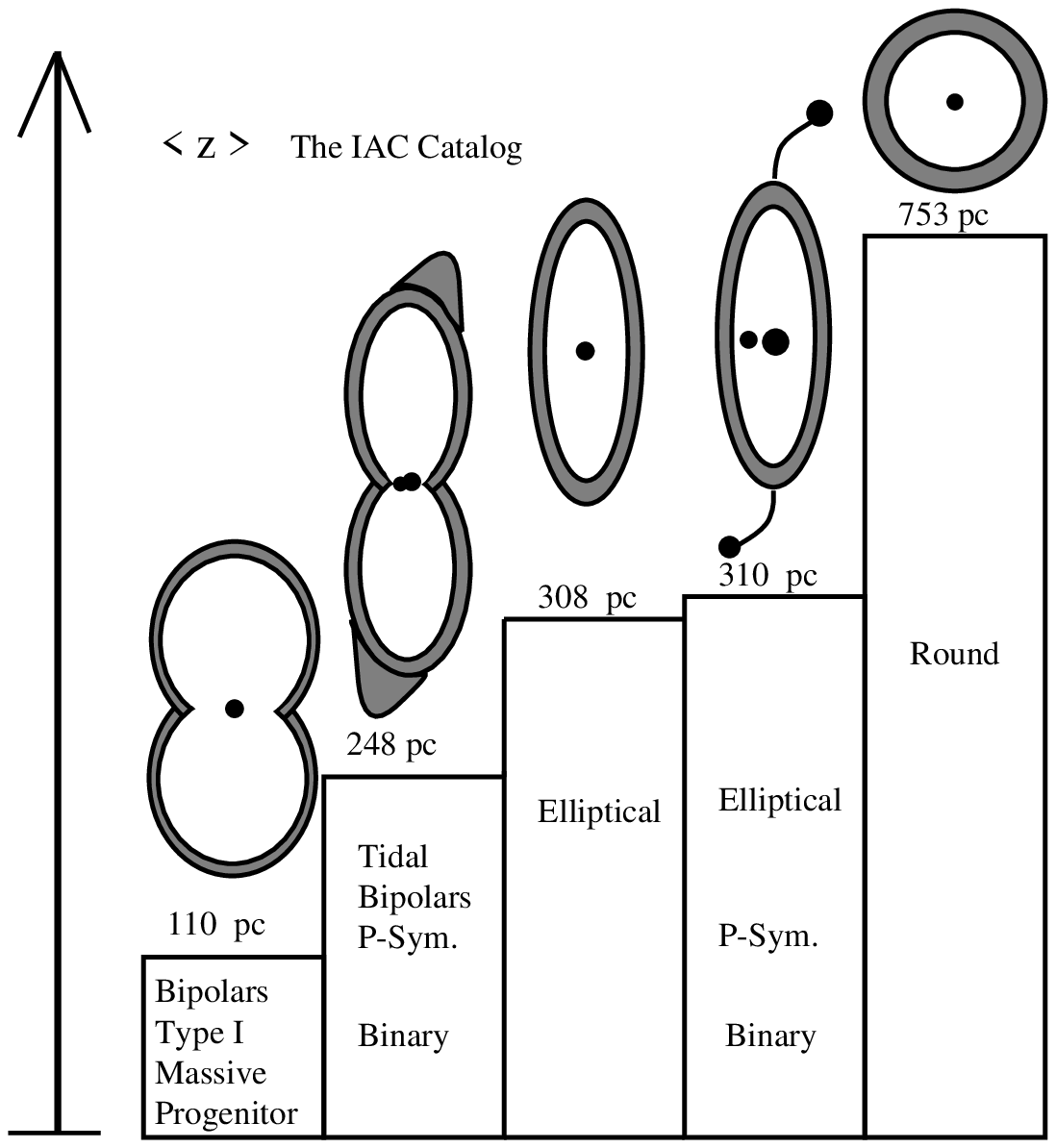}
  \caption{}
  \label{}
\end{figure}

Finally, the relation between morphology and galactic distribution
could be explained as (see Figure 1):

\begin{description}

\item $\bullet$ Bipolars Type I Peimbert = (B):

Small $< z >$  (110 pc) $\Longleftrightarrow$ Massive Progenitor 
$\Longleftrightarrow$ Stellar Rotation $\Longleftrightarrow$
$\Omega$ Limit $\Longleftrightarrow$ Classical Bipolarity 

\item $\bullet$ Tidal Bipolars = (BPS):

Moderate $< z >$  (248 pc) $\Longleftrightarrow$ Non-Massive Progenitor in
Close Binary System $\Longleftrightarrow$ Tidal Spin-Up $\Longleftrightarrow$
Shaping by $\Omega$ Limit + MHD Effects + Precession/Wobbling
 $\Longleftrightarrow$ Bipolarity with Point-Symmetry (Lobes, FLIERS, Jets)

\item $\bullet$ Ellipticals = (E):

Medium $< z >$  (308 pc) $\Longleftrightarrow$ Non-Massive Progenitor
$\Longleftrightarrow$ Shaping by MHD Effects $\Longleftrightarrow$
FLIERS \& Jets with Axisymmetry

\item $\bullet$ Ellipticals with Point-Symmetry = (EPS): 

Medium $< z >$  (310 pc)  $\Longleftrightarrow$ Non-Massive Progenitor in
Wide Binary System $\Longleftrightarrow$ Shaping by MHD Effects + Precession
$\Longleftrightarrow$ FLIERS \& Jets with Point-Symmetry

\item $\bullet$ Round = (R):

Large $< z >$  (753 pc) $\Longleftrightarrow$ Low-mass Progenitor
$\Longleftrightarrow$ Neither Rotation nor MHD effects

\end{description}

\acknowledgements 
G.G.-S. is very grateful to
M. Peimbert, S. Torres-Peimbert, A. Manchado, M.-M. Mac Low, 
Y. Terzian, N. Soker, M. Reyes-Ruiz and
L. Stanghellini for very fruitful discussions and comments. 
We specially thank Michael L. Norman and the Laboratory for 
Computational Astrophysics for the permission to use ZEUS-3D.
The computations were performed at the Instituto de Astronom\'{\i}a-UNAM.
This work has been supported by grants from DGAPA-UNAM (IN 114199) 
and  CONACyT (32214-E).

\begin{description}

\item Balick, B. 1987, AJ, 94, 671

\item Bjorkman, J. E., \& Cassinelli, J. P. 1993, ApJ, 409, 429

\item Blackman, E. G., Frank, A., \& Welch, C. 2001, ApJ, 546, 288

\item Calvet, N. \& Peimbert, M. 1983, RMxAA, 5, 319

\item Contopoulos, J. 1995, ApJ, 450, 616  

\item Corradi, R. L. M. 2000,  
{\em Asymmetrical Planetary Nebulae II: From Origins to Microstructures},
eds. Joel H. Kastner, Noam Soker, Saul A. Rappaport, A.S.P. Conference Series,
199, 25 

\item Corradi, R. L. M., \& Schwarz, H. E. 1995, A\&A, 293, 871 

\item Chevalier, R. A., \& Luo, D. 1994, ApJ, 421, 225

\item Chu, Y.-H., Jacoby, G. H. \& Arendt, R. 1987, ApJS, 64, 529

\item Franco, J. et al. 2001 , RMxAASC , in press

\item Friend, D. B., \& Abbott, D. C., 1986, ApJ 311, 701 

\item Garc\'{\i}a-Segura, G. 1997, ApJL, 489, L189 

\item Garc\'{\i}a-Segura, G.,  Franco, J., L\'opez, J. A., Langer, N., 
\& R\'o\.zyczka, M. 2000, {\em
Asymmetrical Planetary Nebulae II: From Origins to Microstructures}, 
eds. Joel H. Kastner, Noam Soker, Saul A. Rappaport, A.S.P. Conference Series,
199, 235 

\item Garc\'{\i}a-Segura, G., Langer, N., R\'o\.zyczka, M., \&
Franco, J. 1999, ApJ, 517, 767

\item Garc\'{\i}a-Segura, G., \& L\'opez, J. A. 2000, ApJ, 544, 336

\item Greig, W. E. 1972, A\&A, 18, 70

\item Guerrero, M.A., V\'azquez, R., \& L\'opez, J.A. 1998, ApJ, 117, 967

\item Heger, A.,\& Langer, N. 1998, A\&A, 334, 210

\item Ignace, R., Cassinelli, C. P., \& Bjorkman, J. E. 1998, ApJ, 505, 910 

\item Langer, N. 1997, in: {\em Luminous Blue Variables: Massive Stars 
in Transition}, ASP Conf. Ser. Vol.~120, 
San Fransisco, eds.  A. Nota \& H.J.G.L.M. Lamers, p.~83

\item Langer, N., Garc\'{\i}a-Segura, G. \& Mac Low, M.-M. 1999a, 
ApJL, 520, L49

\item Langer, N., Heger A., Wellstein S., Herwig F. 1999b, A\& A, 346, L37

\item Livio, M., \& Pringle, J. E. 1996, ApJL, 465, 55

\item L\'opez, J.A., Meaburn, J. \& Palmer, J.  1993, ApJ, 415, L135

\item Manchado, A., Guerrero, M., Stanghellini, L., \&
Serra-Ricart, M. 1996, 
``The IAC Morphological Catalog of Northern Galactic Planetary
Nebulae'',
ed. Instituto de Astrof\'{\i}sica de Canarias

\item Manchado, A., Villaver, E., Stanghellini, L., \& Guerrero, M. A.
2000, {\em Asymmetrical Planetary Nebulae II: From Origins to Microstructures},
eds. Joel H. Kastner, Noam Soker, Saul A. Rappaport, A.S.P. Conference Series,
199, 17

\item Mastrodemos, N., \& Morris, M. 1998, ApJ, 497, 303

\item Matt, S., Balick, B., Winglee, R., \& Goodson, A. 2000, ApJ, 
545, 965M

\item Miller, G. E., \& Scalo, J. M. 1979, ApJS, 41, 513

\item Nota, A., \& Clampin, M. 1997, in: 
{\em Luminous Blue Variables: Massive Stars
in Transition}, ASP Conf. Ser. Vol.~120,
San Fransisco, eds.  A. Nota \& H.J.G.L.M. Lamers, p.~303

\item Peimbert, M., \& Torres-Peimbert, S. 1971, ApJ, 168, 413

\item Peimbert, M. 1978, in IAU Symp. No 76, 215

\item Peimbert, M., \& Torres-Peimbert, S. 1983, in IAU Symp. No 103, 233

\item Reid, M. J., Moran, J. M., Leach, R. W., Ball, J. A., Johnston,
K. J., Spencer, J. H., \& Swenson, G. W. 1979, ApJL, 227, L89

\item Reyes-Ruiz, M., \& L\'opez, J. A. 1998, ApJ, 524, 952

\item R\'o\.zyczka, M. \& Franco, J. 1996, ApJL, 469, L127

\item Schmidt, G. 1989, in IAU Coll. 114, ``White Dwarfs", ed. G. Wegner 
(Dordrecht, Kluwer), 305

\item Schwarz, H. E., Corradi, R. L. M. \& Melnick, J. 1992, 
             A\&A Suppl., 96, 23

\item Soker, N. 1995, M.N.R.A.S., 274, 147

\item Soker, N.,\& Rappaport, S. 2000, ApJ, 538, 241S

\item Stanghellini, L., Corradi, R. L. M. \& Schwarz, H. E. 1993, 
            A\&A, 276, 463

\item Terzian, Y. 1989, in IAU Symp No 131, Planetary Nebulae, 17

\item Torres-Peimbert, S., \& Peimbert, M. 1977, RMxAA, 2, 181

\item Torres-Peimbert, S., \& Peimbert, M. 1979, RMxAA, 4, 341

\item Torres-Peimbert, S., \& Peimbert, M. 1983, PASP, 95, 601 

\item Torres-Peimbert, S., \& Peimbert, M. 1997, in IAU Symp. No 180, 175

\item Zukerman, B., \& Aller, L. H. 1986, ApJ, 301, 772

\item Zukerman, B., \& Gatley, I. 1988, ApJ, 324, 501 

\end{description}

\end{document}